\documentstyle[12pt,aasms4]{article}
\received{18 March 1997}
\accepted{26 May 1997}

\slugcomment{To appear in Astrophysical Journal Letters}

\lefthead{Pedersen et al.}
\righthead{A New measurement of baryonic fraction}

\begin{document}

\title{A new measurement of the baryonic fraction\\
using the sparse NGC 3258 group of galaxies}

\author{Kristian Pedersen\altaffilmark{1}}
\affil{Niels Bohr Institute, Blegdamsvej 17, DK-2100 Copenhagen, Denmark; 
kristian@dsri.dk}

\author{Yuzuru Yoshii}
\affil{Institute of Astronomy, University of Tokyo,
Mitaka, Tokyo 181, Japan; yoshii@omega.mtk.ioa.s.u-tokyo.ac}
\affil{Research Centre for the Early Universe, University
of Tokyo, Bunkyo-ku, Tokyo 113, Japan}

\and

\author{Jesper Sommer--Larsen}
\affil{Theoretical Astrophysics Centre, Juliane Maries Vej 30,
DK-2100 Copenhagen, Denmark; jslarsen@tac.dk}

\altaffiltext{1}{Present address: Danish Space Research Institute,
Juliane Maries Vej 30, DK-2100 Copenhagen, Denmark} 

\begin{abstract}
New X-ray observations of the sparse NGC 3258 group of galaxies made by the 
ASCA satellite with good spectral and spatial resolution has revealed that this
group has a gravitational potential deep enough to prohibit significant 
mass removal from the system. The baryonic fraction within 
$240h_{50}^{-1}$~kpc is found to be $0.065^{+0.051}_{-0.020}$ for 
$h_{50}=1$, where $h_{50}\equiv H_0/50$~km$\,$s$^{-1}$Mpc$^{-1}$, 
in good agreement with the universal value of $0.05\pm 0.01$ predicted by 
standard Big Bang nucleosynthesis for a Universe with $\Omega_0=1$ and 
$h_{50}=1$.
Since the deep potential of the NGC 3258 group ensures that all pristine
intragroup gas has been retained, the baryonic fraction of the NGC 3258 
group is indicative of the universal value. Consequently it seems premature 
to rule out a critical Universe.
\end{abstract}


\section{Introduction}

A key issue in cosmology is to determine the baryonic fraction in the
Universe and X-ray observations are particularly well suited for this task.
They have revealed that the main component
of light-emitting matter in clusters of galaxies is a $T\sim 10^7$ degree 
hot gas trapped in the cluster potential (\cite{white}, \cite{david}).
Recent X-ray studies with ROSAT have shown that the most abundant galaxy 
associations (\cite{tully}, \cite{nolt}) groups of galaxies with only a 
handful of galaxies, are also massive enough to retain hot intergalactic 
gas (\cite{pon}, \cite{mul93}). The X-ray emitting hot gas seems
to be rather smoothly and spherically distributed, lending support to the 
assumption that the gas is in hydrostatic equilibrium with the underlying 
potential. With this assumption, and since the hot gas is detected far 
beyond the optical extent of the group, the gravitating mass distribution 
can be determined to several hundred kiloparsecs from the group center.
The hot gas in groups is therefore a much more secure tracer of the 
gravitational field than the galaxies which do not even assure the 
physical reality of their grouping on the sky.\\ 

The mass fraction of baryons consisting of light-emitting matter (galaxies 
and hot gas) in the few groups with well-determined masses is around 
$f_b\approx 0.15$ for $h_{50}=1$ (\cite{mul96}) which is not too dissimilar to 
$f_b \simeq 0.20\! -\! 0.25\,h_{50}^{-3/2}$ (\cite{white}, 
\cite{david}) in clusters of galaxies. If groups or clusters are
representative of the Universe as a whole, the above baryonic fractions 
may be identified with the ratio of baryon to total mass densities 
$\Omega_b/\Omega_0$ in the Universe. Neither the baryonic fraction of the 
groups studied so far nor the baryonic fraction of clusters can be 
reconciled with the standard Big Bang nucleosynthesis 
(SBBN) prediction $\Omega_bh_{50}^2=0.05\pm 0.01$ for a critical 
$\Omega_0=1$ Universe if $h_{50}=1$ (\cite{simon}).\\ 

One approach for narrowing in on the universal baryonic fraction 
is to obtain deep X-ray observations of galaxy groups with a sufficiently 
deep potential, to ensure that de-gassing has not been important. 
Galaxy groups are much more abundant than galaxy clusters and those
groups having kept all their primordial gas should give 
a relatively fair estimate of the baryonic fraction in the Universe.\\

The Advanced Satellite for Cosmology and Astrophysics (ASCA) provides 
superior energy coverage and spectral resolution compared to previous X-ray 
observatories, yielding more accurate determination of temperature and 
metal abundance of hot intergalactic gas in galaxy associations.  
In order to study the hot gas content of a deep potential group and 
determine its content of light-emitting matter and dark matter, 
we have obtained new ASCA observations of the NGC 3258 group for which 
the optical parameters are listed in Table~1. The four mass estimators 
of \cite{heisler} yield masses in the range 
$0.9 - 2.4\times 10^{13}M_{\odot}$ with a mean of 
$M_{vir}=1.7\times 10^{13}M_{\odot}$ being uncertain within a factor 
of about two. The large virial mass of the NGC 3258 group indicates 
that it has a deep gravitational potential. Furthermore the short crossing
time and short collapse time point to a bound and relaxed system.

\placetable{table1}

\section{X-ray studies of the NGC 3258 group of galaxies}

We obtained a 22400 sec exposure of the NGC 3258 group of galaxies 
with ASCA in May 1994.
More than 4000 photons from an extended source centered on the NGC 3258 
group were detected with each of the two Gas Imaging Spectrometers (GISs).
One of the Solid-state Imaging Spectrometers (SIS0) detected more than
1800 photons whereas the other (SIS1) were so badly affected by ``hot and
flickering'' pixels that the NGC 3258 group emission from this detector 
could not be recovered.  Consequently the SIS1 data were excluded from 
the analysis. The ASCA data were screened for local contaminating sources 
and background emissions from the detectors, and the sky were subtracted
using the publically available long exposures at high Galactic latitude.

\placefigure{g2}

Diffuse emission centered on the group is clearly detected by each of 
the GIS detectors and by the SIS0 detector (Fig.~\ref{g2}). In the SIS0 image
there is a clear peak situated at the centroid of the diffuse emission.
Furthermore a point-like source North-West of the group center is detected 
and diffuse emission is extending North-East-wards beyond the GIS field of 
view, towards the poor cluster around NGC 3268.

The point-like source is $11^{\prime\prime}$ off-set a background Seyfert 
galaxy located at a redshift of $z=0.0636$ 
(M.I.\ Andersen, private communication). 
The X-ray emission is well fitted as a point source shining through the 
intragroup gas. No other obvious optical counterparts are found within the 
$40^{\prime\prime}$ pointing error circle (90\% confidence) of ASCA. 
In the analysis of the diffuse group emission a circular region centered
on the X-ray point source was masked out. For the GIS data, a radius of
$4^{\prime}$ circle was chosen, and for the SIS0 data a radius of
$3^{\prime}$.

In order not to be biased by the emission extending towards the NGC 3268 
cluster, only photons within $10^{\prime}$ of the optical axis in this 
quadrant of the detector were included in the model fits to the GIS images.
The contribution of ``stray photons'' from the NGC 3258 cluster outside the 
field of view were estimated from ray tracing simulations, kindly provided 
by the astrophysics group of Nagoya University.  At most, a few percent of 
the photons within any region in the GIS field of view originates outside 
the field of view.

The peak of the diffuse emission is off-set NGC 3258 by $1.1^{\prime}$ and 
off-set NGC 3260 by $1.8^{\prime}$, making it most likely that the peak is 
attributed to the minimum of the group potential.  
We analysed the surface brightness distribution of the diffuse intragroup 
emission using a model-fitting procedure (\cite{pede}). 
Given the spherically symmetric appearance of the group X-ray emission and 
the number of detected photons, we generated azimuthally symmetric model 
images of the group emission characterised by the so-called $\beta$-model, 
$S(r)=S_0(1+(r/r_c))^{-3\beta/2}$, generally used to fit the X-ray surface 
brightness profile of clusters of galaxies (\cite{jf}). 
Vignetting of the X-ray telescopes was applied and the model images were 
folded with the point spread function of the X-ray telescopes and the GIS 
detectors.  We found from the spectra that most of the photons were in the 
range of 1--2~keV, so that we adopted the vignetting and point spread 
functions at an energy of 2~keV.  We checked that using the telescope and 
detector responses at an energy by 1 keV lower or higher does not change 
the conclusion of our analysis.  Model images were then fitted individually 
to each GIS image and the SIS0 image with the slope of the surface brightness
profile and the core radius as free parameters.  Due to the low number of 
counts per pixel element a maximum-likelihood statistic was minimised 
rather than $\chi^2$.

Model fits to each of the GIS images were mutually consistent and 
simultaneous fit to both GIS images yielded a slope of  $\beta=0.6\pm0.2$ 
and a core radius of $r_c=112\pm50\,h_{50}^{-1}$~kpc.  These two parameters 
are tightly linked in a way that larger $\beta$ results in larger $r_c$.  
The quoted uncertainties arise from varying the level of the background 
emission by 20\% which dominates over the statistical errors for a low 
surface brightness source like the NGC 3258 group.  The intragroup emission 
is detected out to at least $240\,h_{50}^{-1}$~kpc.

The diffuse emission extends all the way to the edge of the SIS0 field of 
view, making it difficult to constrain the surface brightness profile from 
this detector.  However, the best-fit profile from the SIS0 image is 
entirely consistent with the best-fit profile from simultaneous fit to 
both GIS images.  The three independent ASCA images give tight constraints 
on the distribution of intragroup gas in the NGC 3258 group. 

\placefigure{spec}

Integrated spectra of the masked region within $15^{\prime}$ of the centroid
of the group peak for the GIS detectors and within $14^{\prime}$ for the 
SIS0 detector were accumulated.  The Fe-L complex around 1 keV, which is 
typical of a 1--2 keV plasma, is visible in the SIS0 spectrum 
(Fig.~\ref{spec}).
This enables reliable determination of metal abundance in the intragroup 
gas.  Plasma emission models were folded with the instrument responses 
and fitted to the spectra, using the XSPEC version 9.00 software. 
MEKAL models (\cite{mekal}) as well as Raymond-Smith models (\cite{rs})
provide good 
fits to all three spectra.  Simultaneous fit of MEKAL models to the GIS2 
and the GIS3 spectra yields a temperature of $kT=1.90^{+0.21}_{-0.18}$ keV 
and a metal abundance of $Z=0.11^{+0.15}_{-0.10}$ solar (1 sigma errors) 
with $\chi^2=117$ for 141 degrees of freedom.  Best-fit values of these
parameters for the SIS0 spectrum are $kT=1.58^{+0.16}_{-0.13}$~keV and 
$Z=0.09^{+0.06}_{-0.05}$ solar ($\chi^2=59.3$ for 61 degrees of freedom). 
Fully consistent results are obtained when Raymond-Smith models are used.  
While the metal abundance of the NGC 3258 group is very low in concordance 
with the abundances of other groups deduced from lower resolution ROSAT 
spectra (\cite{mul96}), the temperature $kT=1.7\pm 0.2$~keV is significantly 
higher than those in the few other groups where the intragroup gas has been 
detected (\cite{mul96}).  We also extracted spectra from three annuli 
centered on the 
centroid of the diffuse emission.  In the process of spectral fitting, we 
took the mixing of photons between annuli into account, but for the three 
wide annuli chosen a maximum of 25\% of the photons originated outside the 
annulus itself.  No significant variations of $kT$ and $Z$ were seen out 
to $15^{\prime}$.

\section{Mass determinations of the NGC 3258 group}

From the profiles of gas density and temperature, we estimated the 
intragroup gas mass and the gravitating mass in the NGC 3258 group.   
Integrating the best-fit $\beta$ model out to the detected extent, 
the gas mass inside $240 h_{50}^{-1}$~kpc is directly obtained as 
$M_{gas}=7.8\times 10^{11} h_{50}^{-5/2}M_{\odot}$.   On the other hand,
the mass of galaxies is $M_{gal}=7.1\times 10^{11}h_{50}^{-1}M_{\odot}$,
assuming a standard mass to light ratio of $M/L_B=8h_{50}(M/L_{B})_{\odot}$
for NGC 3258 and NGC 3260 and $M/L_B=3 h_{50}(M/L_{B})_{\odot}$ for NGC 3257.

The relatively high temperature and steep density profile of the intragroup 
gas implies that 
the gravitational potential well is deeper than in the few other sparse groups
which have been subject to detailed X-rays studies. It should be safe 
to assume that no gas has been expelled from the group by galactic winds 
(\cite{ds}, \cite{ya}). The smooth 
appearance and short sound crossing time of the intragroup gas ensure that the 
gas is approximately in hydrostatic equilibrium with the group potential, 
allowing us to estimate the gravitating mass.  In concordance with the 
temperature profile deduced from the spectral fits, we model the intragroup 
gas as isothermal gas, giving the following expression for the 
gravitating mass (\cite{cowie87})
\begin{equation}
    M_{grav}(r)=1.8\times 10^{12}h_{50}^{-1}M_{\odot}\,
    (kT/1\,{\rm keV})
    (r/1^{\prime})\; 
    \beta\;\frac{(r/r_c)^2}{1+(r/r_c)^2} \;\;\; .
\end{equation}
Numerical simulations of clusters of galaxies show that this mass estimator
is unbiased and accurate to better than 30\%, even when significant 
substructure is present (\cite{evrard}).  Using $\beta=0.6$, 
$r_c=112 h_{50}^{-1}$~kpc, and $kT=1.7$~keV, the gravitating mass inside 
$240 h_{50}^{-1}$~kpc is obtained as 
$M_{grav}=2.3\times 10^{13}h_{50}^{-1}M_{\odot}$.
Since $M_{grav}$ far exceeds $M_{gas}$ and $M_{gal}$, we conclude that most
of mass in the NGC 3258 group is dark which is reflected in a large 
mass to light ratio $M_{grav}/L_B=237\,h_{50}(M/L_B)_{\odot}$.
The virial mass estimate is a bit lower than the X--ray determined mass.
This is to be expected since the galaxy motions only probe the inner parts
of the group whereas the X--ray data trace
mass contributions out to $240h_{50}^{-1}$~kpc from the group center.\\

The baryonic fraction or the mass ratio of galaxies and gas
to the total gravitating matter inside $240h_{50}^{-1}$~kpc is given by
\begin{equation}
  f_b=\frac{M_{gal}+M_{gas}}{M_{grav}}=0.031 + 0.034\,h_{50}^{-3/2}\;\;\;.
\end{equation}
The baryonic fraction at large radii is nearly independent of the radius at 
which it is evaluated (Fig.~\ref{fbar}), decreasing from 
$0.092^{+0.041}_{-0.016}$ ($h_{50}=1$) at $r=150 h_{50}^{-1}$~kpc to 
$0.065^{+0.051}_{-0.020}$ ($h_{50}=1$) at $r=240 h_{50}^{-1}$~kpc.

\placefigure{fbar}

The low baryonic fraction relative to the few other groups with 
well-determined gas fractions is due to a relatively steep gas density profile
and a high intragroup gas temperature, yielding a low gas mass and a high 
gravitating mass.  A summary of the results is given in Table~2.

\placetable{table2}

\section*{\large\bf Cosmological implications}
Our deep ASCA pointing of the sparse NGC 3258 group has enabled a precise 
determination of the baryonic fraction in this system. The derived 
value of $f_b\approx 0.065$ is comparable to the SBBN 
prediction of the universal baryonic density 
$\Omega_bh_{50}^2=0.05\pm 0.01$ with $\Omega_0=1$ and $h_{50}\approx 1$.

Our discovery of a sparse galaxy group having a low baryonic fraction 
is interesting to discuss in light of the cosmological implications 
associated with recent measurements of primordial deuterium in high-redshift 
quasar absorption line systems (\cite{rh}, \cite{tytler}, \cite{songaila}). 
Although the reported $D/H$ abundance is still 
controversial, a range of $4\times 10^{-5}< D/H <2.4\times 10^{-4}$ seems 
to persist (\cite{songaila}), yielding a universal baryonic density of 
$0.022<\Omega_b\,h_{50}^2<0.064$ from SBBN theory (\cite{hogan}) where
the higher $D/H$ corresponds to the lower $\Omega_b$.
If the baryonic fraction derived in clusters (\cite{white}, \cite{david}) of 
$f_b \simeq 0.20\! -\! 0.25\,h_{50}^{-3/2}$ is representative of the Universe, 
the total matter density in the Universe would be 
$0.09<\Omega_0\,h_{50}^{-1/2}<0.32$. However, taking the baryonic fraction of 
the NGC 3258 group as representative, motivated by the ability of its 
deep potential to retain pristine gas, the estimated total matter density is
$0.34^{+0.15}_{-0.16}<\Omega_0<0.99^{+0.44}_{-0.45}$ ($h_{50}=1$) or
$0.20^{+0.07}_{-0.05}<\Omega_0<0.57^{+0.17}_{-0.20}$ ($h_{50}=1.5$).
We note in passing, that these bounds are in accord with constraints on 
$\Omega_0$ from large-scale flows (\cite{dekel}) and from the first 
acoustic peak in the cosmic microwave background (\cite{hancock}).

In conclusion, our measured low $f_b$ value removes much of the prevailing 
argument that the baryonic fraction in galaxy systems only leaves room for a 
low-density Universe. Specifically, it is premature to rule out a critical 
Universe, and we conjecture that such a Universe is still a 
viable option in modern cosmology.\\

\acknowledgments

We are grateful to the ASCA Team for assistance
with the analysis of ASCA data. Y.\ Tawara, M.\ Watanabe, A.\ Furuzawa and
F.\ Akimoto of Nagoya University are gratefully acknowledged for providing 
ray tracing simulations of the stray light effects from the NGC 3268 cluster.
We thank Hans Ulrik N\o rgaard--Nielsen for enlightning discussions and 
useful suggestions to the manuscript.  This work has been supported in part
by the Grant-in-Aid for Center-of-Exellence Research (07CE2002) of the
Ministry of Education, Science, and Culture, Japan.

\clearpage

\begin{table*}
\begin{center}
{\bf TABLE 1}\\
{\bf NGC 3258 GROUP PROPERTIES}

\vspace{0.5cm}

\begin{tabular}{cccc}
\tableline \tableline
&&&\\
\multicolumn{4}{c}{Parameters for individual member galaxies$\;^a$} \\
\multicolumn{4}{c}{\rule{9.5cm}{0.2mm}}\\
&& $L_B$ & $v_r$ \\
Object & Type & $(10^{10}\,h_{50}^{-2}\,L_{B\odot})$ & (km$\,$s$^{-1}$)\\
\tableline
NGC 3258 &\multicolumn{1}{l}{E1}     & 6.2 & 2792 \\
NGC 3260 &\multicolumn{1}{l}{E$_p$}  & 2.1 & 2416 \\
NGC 3257 &\multicolumn{1}{l}{SAB(s)0}& 1.4 & 3172 \\
\tableline
\end{tabular}
\begin{tabular}{ccccc}
&&&&\\
\multicolumn{5}{c}{Parameters for the NGC 3258 group$\;^b$}\\
\multicolumn{5}{c}{\rule{9.5cm}{0.2mm}}\\
$\sigma$ & $t_{cross}$ & $t_{coll}$ & $M_{vir}$ & $M_{gal}$ \\
(km$\,$s$^{-1}$) & $(H_0^{-1})$ & $(H_0^{-1})$ & 
$(10^{12}\,h_{50}^{-1}\, M_{\odot})$ & 
$(10^{12}\,h_{50}^{-1}\, M_{\odot})$ \\
\tableline
400 & 0.07 & 0.13 & 17 & 0.71 \\
\tableline
\end{tabular}
\end{center}

\tablenotetext{a}{Galaxy type, blue luminosity $L_B$, heliocentric radial 
velocity $v_r$ are drawn from RC3.  The blue luminosity $L_B$ has been 
corrected for internal and Galactic absorption assuming a distance of 
$56\,h_{50}^{-1}$~Mpc.} 
\tablenotetext{b}{Galaxy positions are drawn from RC3. The velocity 
dispersion $\sigma$ has been corrected for measurement errors. 
The crossing time $t_{cross}$ is evaluated as the 
inertial radius (\cite{jackson}) divided by the velocity dispersion.  
The collapse time $t_{coll}$ is estimated assuming a homogenous sphere 
with the group virial mass $M_{vir}$ which initially expands with the 
universe and then collapses under its own gravity (\cite{gg}).  The mass of 
galaxies $M_{gal}$ is estimated from their blue luminosities assuming 
$M/L_B=8\, h_{50} (M/L_{B})_{\odot}$ for NGC 3258 and NGC 3260 and 
$M/L_B=3\, h_{50} (M/L_{B})_{\odot}$ for NGC 3257.}

\caption{
\label{table1}}
\end{table*}

\clearpage

\begin{table*}
\begin{center}
{\bf TABLE 2}\\
{\bf NGC 3258 GROUP X-RAY RESULTS$\;^a$}

\vspace{0.5cm}

\begin{tabular}{ccccccc}
\tableline \tableline
& $r_c$ & $kT$ & $Z$ & $L_X$ & $M_{gas}$ & $M_{grav}$ \\
$\beta$ & $(h_{50}^{-1}$~kpc) & (keV) & $(Z_{\odot}$) & 
$(10^{42}\,h_{50}^{-2}$~erg/s) & $(10^{12}\,h_{50}^{-5/2}\,M_{\odot})$ &
$(10^{12}\,h_{50}^{-5/2}\,M_{\odot})$ \\
\tableline
$0.6\pm 0.2$ & $110\pm 50$ & $1.7\pm 0.2$ & $0.1\pm 0.1$ & $4.3\pm 0.3$ &
$0.78^{+0.32}_{-0.24}$ & $23^{+6}_{-8}$ \\
\tableline
\end{tabular}
\end{center}

\noindent

Note -- Quoted errors on derived X-ray parameters are those from the 
systematics which dominate over the statistical errors.

\tablenotetext{a}{The temperature $kT$ and metal abundance $Z$ are the 
best-fit values when combining integrated spectra within $14^{\prime}$ 
of the group center for SIS0 and within $15^{\prime}$ of the group center 
for GIS2 and GIS3.  The X-ray luminosity $L_X$ in the 0.5--5 keV band is 
estimated from the half of intragroup emission which is not contaminated 
with the background Seyfert emission or the extension towards NGC 3268
assuming a spherically symmetric surface brightness distribution.
The gas mass $M_{gas}$ and the gravitating mass $M_{grav}$ are those 
integrated out to $r=240\, h_{50}^{-1}$~kpc from the group center assuming 
that the intragroup gas is isothermal and in hydrostatic equilibrium.}
\caption{
\label{table2}}
\end{table*}

\clearpage

\figcaption[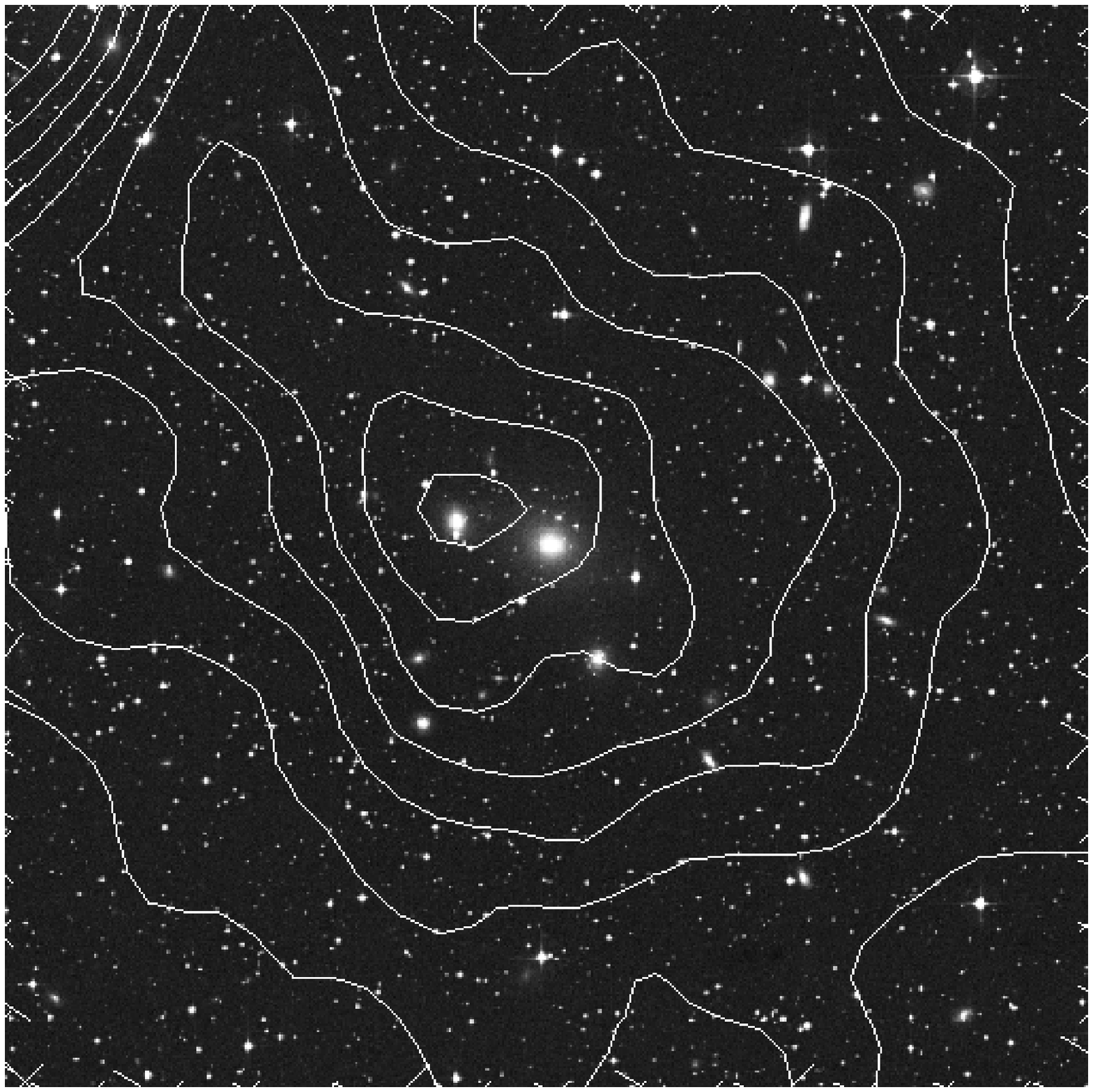]{Optical image from the Digitized Sky Survey over-layed with
ASCA GIS2 contours. The GIS2 contours were generated from the background
subtracted image smoothed to $2^{\prime}$ resolution.  A point-source model
of the emission from the background Seyfert was subtracted. 
No correction for vignetting was applied to the image. 
The field is $30^{\prime}$ by 
$30^{\prime}$ with North up and East to the left. The contour scale is linear, 
ranging from $8.9\times 10^{-5}$~counts/arcmin$^2$/sec to 
$7.6\times 10^{-4}$~counts/arcmin$^2$/sec. In the North-Eastern corner,
contour lines are crowding at the edge of the detector boundary.
\label{g2}}

\figcaption[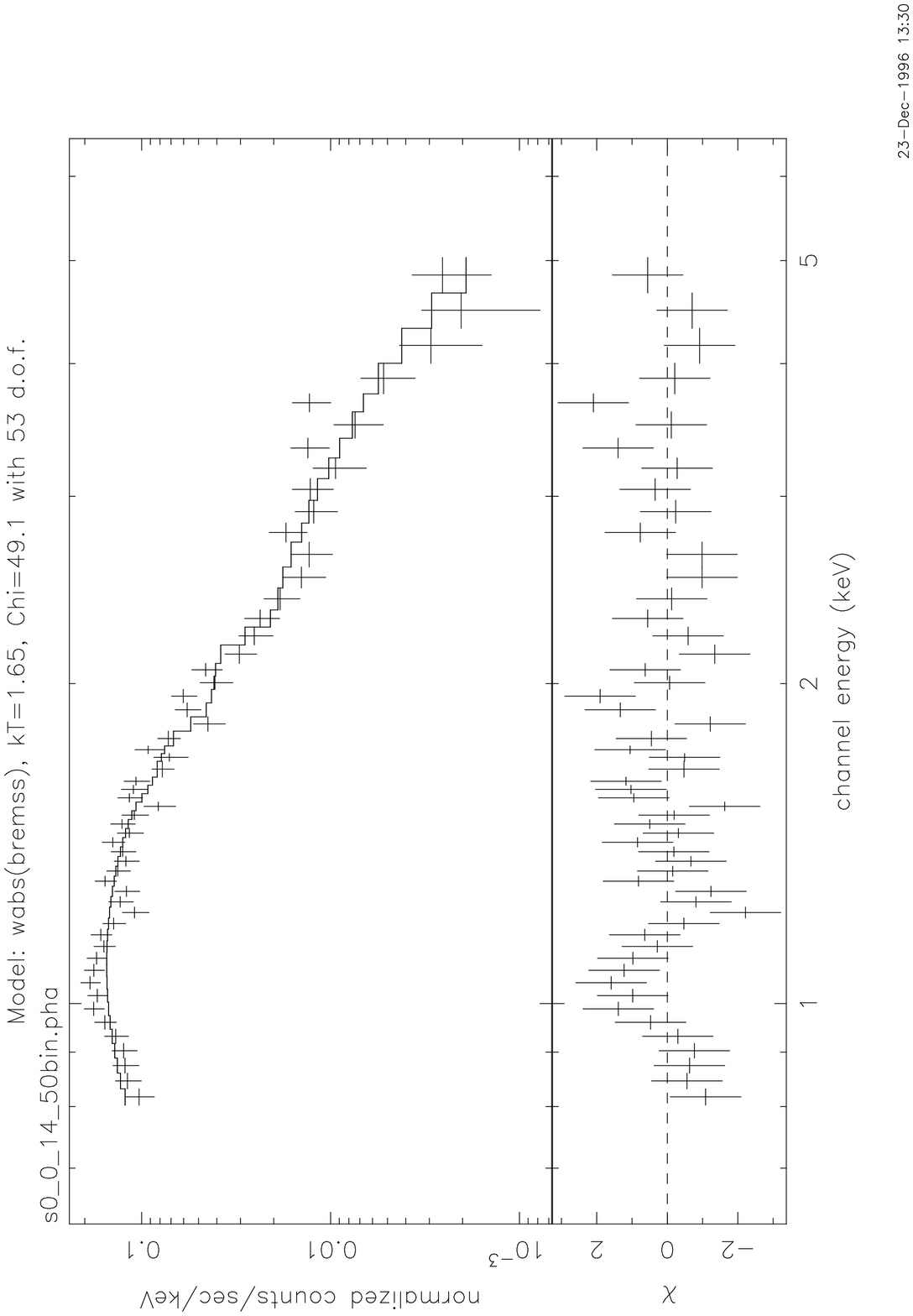]{Integrated SIS0 spectrum within $14^{\prime}$ of 
the group center. Detections from a $3^{\prime}$ circular region around 
the Seyfert galaxy were excluded. The histogram is the best fit 
bremsstrahlung model with absorption from Galactic hydrogen, 
$kT=1.65^{+0.16}_{-0.15}$~keV. Below the residuals are shown in units of 
standard deviations. The model is acceptable ($\chi^2=49$ for 53 degrees of 
freedom), but systematic residuals due to the Fe-L line complex are seen
around 1~keV.
\label{spec}}

\figcaption[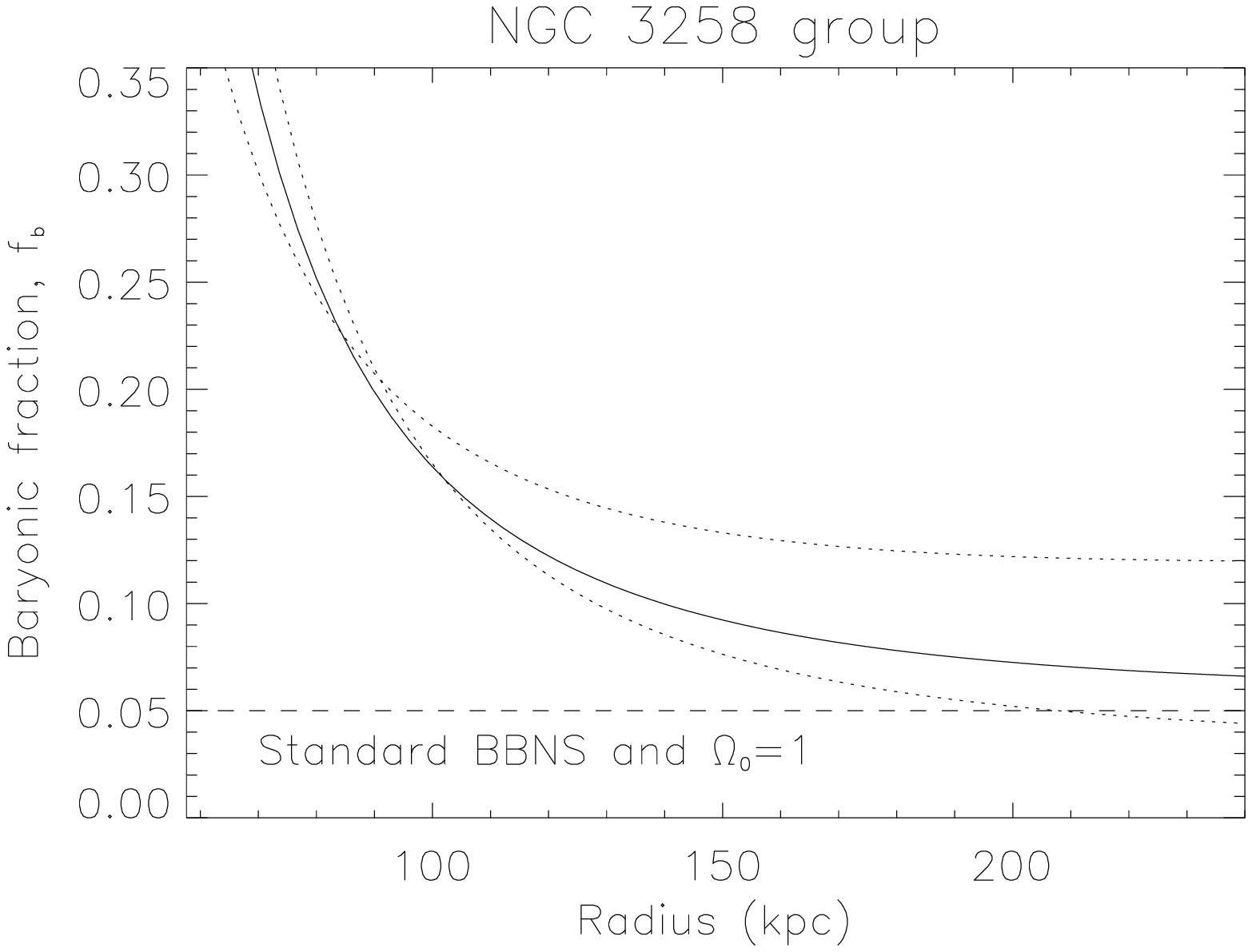]{Baryonic fraction of the NGC 3258 group versus radial 
distance from the group center for $h_{50}=1$. The solid line is the 
baryonic fraction 
from the best fit intragroup gas temperature ($kT=1.7$~keV) and the best fit 
intragroup gas density distribution ($\beta=0.6$, $r_c=112 h_{50}^{-1}$~kpc). 
The dotted lines show extreme values of the baryonic fraction when combining 
errors on the gas temperature with errors on the gas density distribution.
The dashed line marks the standard big bang nucleosynthesis prediction 
for a critical $\Omega_0=1$ Universe (Hogan 95) with $h_{50}=1$.
\label{fbar}}

\clearpage

\plotone{g2.ps}

\clearpage

\plotone{s0spec.ps}

\clearpage

\plotone{fbar.ps}

\clearpage

\end{document}